\documentclass[a4paper,12pt]{article}
\usepackage{mathrsfs}
\usepackage{amsmath}
\usepackage{amscd}
\usepackage{amssymb}
\usepackage{cite}
\usepackage[dvips]{color}
\renewcommand{\theequation}{\arabic {section}.\arabic{equation}}

\begin{document}


\begin{titlepage}
\begin{flushright}
EPHOU-05-009 \\
December
, 2005
\end{flushright}

\vspace{5cm}

\begin{center}
{\Large Topological hypermultiplet on $N$=2 twisted superspace in four dimensions} \\

\vspace{1cm}

{\scshape Junji Kato}\footnote{jkato@particle.sci.hokudai.ac.jp}, 
{\scshape Akiko Miyake}\footnote{miyake@particle.sci.hokudai.ac.jp}\\

{\textit{ Department of Physics, Hokkaido University }}\\
{\textit{ Sapporo, 060-0810, Japan}}\\
\end{center}

\vspace{2cm}

\begin{abstract}
We propose a $N$=2 twisted superspace formalism with a central charge in
 four dimensions by introducing a Dirac-K\"ahler twist. Using this
 formalism, we construct a twisted hypermultiplet action and find an
 explicit form of fermionic scalar, vector and tensor transformations.
We construct an off-shell Donaldson-Witten theory coupled to the twisted
 hypermultiplet. We show that this action possesses $N$=4 twisted
 supersymmetry at the on-shell level. It turns out that the
 four-dimensional Dirac-K\"ahler twist is equivalent to Marcus'
 twist. 
\end{abstract}

\end{titlepage}

\newpage
\renewcommand{\theequation}{\arabic {section}.\arabic{equation}}
\section{Introduction}
In 1988 Witten pointed out that the $N$=2 super-Yang-Mills theory
corresponds to the Donaldson-Witten theory \cite{W} which is a kind of 
topological field theories(TFT). 
This correspondence is called topological twist or simply twist. This
topological theory was soon after derived from quantizing a four
dimensional topological Yang-Mills theory using instanton gauge
fixing \cite{bs,bms,lp,BRT1}.
This suggest that a topological field theory can be constructed by 
twisting an extended supersymmetric gauge theory or by quantizing a
topological invariant with a suitable gauge fixing condition in gauge
theories. 
From the twisting procedure a BRST charge appearing in TFT was
identified with a part of the supercharges appearing in supersymmetry (SUSY). 
Thus in the quantized TFT there should be other BRST-like fermionic
symmetries. 
A vector supersymmetry \cite{BRT,BR}, that is, a BRST-like fermionic
symmetry with a vector index  was actually discovered in various models
and dimensions
\cite{LSSTV,DGS,GMS,BM,MSore,LPS,LSZ,BSSV,GGPS,GGNPS,FTVVSS}.   
It was recognized that the BRST symmetry and the vector SUSY belong to a
twisted version of the $N$=2 or $N$=4 extended SUSY. 
For example in two dimensional $N$=2 case the twisted supercharges
consist of a scalar (BRST), a vector and a pseudo-scalar (the second rank
tensor) charge \cite{LL,CLS}.
One of the important characteristics of a quantization of TFT is that
the twisted supersymmetry (TSUSY) spontaneously appears after a
quantization. 
This suggests that the origin of the SUSY  may be connected with
quantization.

Kawamoto and Tsukioka proposed a new twisting procedure called
Dirac-K\"ahler twist in two dimensional quantized topological Yang-Mills
theory with instanton gauge \cite{KT} constructed from the generalized
gauge theory \cite{KW,KW2,KOS,KSTU}. It was found that the twisting
procedure between the spinors (gaugino and matter field) and the tensor
fermions (ghost, anti-ghost) is essentially the Dirac-K\"ahler fermion
mechanism \cite{IL,Kahler,G,BJ,Rabin,BDH,BennT,Bull} and the flavor
degrees of freedom of Dirac-K\"ahler fermion can be interpreted as that
of the extended SUSY \cite{DVF,SchapT}.
One of the authors (J.K.) with Kawamoto and Uchida pointed out that the
twisted superspace formalism is hidden behind the formulation
\cite{KKU}. 
It became clear that the two dimensional quantized topological Yang-Mills
theory and the quantized BF theory were successfully derived from the
twisted superspace formalism.

In the previous paper \cite{KKM} the authors (J.K. and A.M.) with Kawamoto
proposed a $N$=4 Dirac-K\"ahler twisting procedure and a twisted
superspace formalism in four dimensions. It turned out that a $N$=4
off-shell twisted supersymmetric action which corresponds to the two
dimensional counterpart of BF theory was constructed from this formalism, but the
action has higher derivative terms and many auxiliary fields and
is not actually the four dimensional BF theory.    
In the $N$=2 case the super Yang-Mills theory whose component fields
belong to the vector multiplet was constructed by using a
superconnection formalism \cite{GSW,Sohnius}. However a matter multiplet
or equivalently  hypermultiplet
\cite{Fay,Soh} has not been found because of the necessity for a central charge. 

In this paper we concentrate on studying $N$=2 twisted SUSY with
a central charge. 
We propose a $N$=2 twisted superspace formalism with the central charge
using the Dirac-K\"ahler twist. 
Related works in a similar context to that of our
formulation was given by Alvarez and Labastida for $N$=2 twisted SUSY by
spinor formulation in four dimensions\cite{AL}, while our formulation is
based on the tensor formulation coming from the Dirac-K\"ahler twisting
procedure. 
We then propose a new twisted hypermultiplet action and give a gauge
covariant version of this action. We claim that this action plus
Donaldson-Witten action
 has the on-shell $N$=4 TSUSY and the four-dimensional
Dirac-K\"ahler twist is equivalent to the Marcus's twist \cite{Mac}. 

There is another off-shell formulation which is called harmonic superspace
formulation \cite{Harmonic}. 
It is interesting that this formulation gives all the off-shell $N$=2
supersymmetric theories.  
It is characterized by a presence of an infinite number of auxiliary
fields and unconstrained superfields.  
It is interesting to construct a twisted version of this formulation and
to establish a relation to the superspace formulation presented here.

Another important motivation of this work comes from the recent study of
lattice SUSY. 
It is well known that the Dirac-K\"ahler fermion mechanism is
fundamentally related to a lattice formulation \cite{KSuss,Suss,KS}. 
In the two dimensional case $N$=2 exact supersymmetry is realized on the
lattice by using the superspace formalism based on the Dirac-K\"ahler
twist \cite{DKKN}. 
These types of lattice SUSY models are considered in three and four
dimensions \cite{DKKN2}.  
Other lattice SUSY models using the Dirac-K\"ahler mechanism were
investigated in \cite{C}.
There are other models preserving part of SUSY on the lattice
based on the TSUSY \cite{C2,S}.

The paper is organized as follows.
In section 2 we give a general property of the $N$=2 twisted superspace
formalism with a central charge coming from the Dirac-K\"ahler twisting
procedure. Then we propose a new topological hypermultiplet action
without interactions.
In section 3 we introduce a Donaldson-Witten theory coupling to the
hypermultiplet. We show that this action possess a $N$=4 TSUSY at on-shell
level and the four-dimensional Dirac-K\"ahler twist is equivalent to the
Marcus's twist. 
In section 4 we establish a connection to the ordinary $N$=4
supersymmetric theory.  
We summarize the result in section 5.
We provide several appendixes to summarize the notations and show the
full transformation of on-shell $N$=4 TSUSY.

\renewcommand{\theequation}{\arabic {section}.\arabic{equation}}
\section{Twisted SUSY with Central Charge}
We have investigated the properties of the four dimensional $N$=4 and $N$=2
twisted superspace formalism without a central charge in a previous
paper \cite{KKM}. In the $N$=2 case the Donaldson-Witten theory was
constructed in our superspace formulation. 
It is well known that another multiplet exists in
ordinary $N$=2 supersymmetric theories. Since this multiplet, that is,
a hypermultiplet accompanies the central charge at off-shell level,
it is difficult to deal with such a vector multiplet.
 
We consider a topological version of the hypermultiplet.
The hypermultiplet satisfies the following $N$=2 SUSY algebra
with the central charge, 			 
\begin{eqnarray}
\{Q _{\alpha i}, \overline{Q}^j _\beta \} = 2 \delta_i ^{j} (\gamma^\mu)_{\alpha\beta} P_\mu +2 \delta_i ^{j} \delta_{\alpha \beta}Z,
\label{eq:SUSY al. with central charge}
\end{eqnarray}
where the indices $\{\alpha,\beta\}$ and the indices $\{i,j\}$ are
Lorentz spinor and internal R-symmetry indices of the extended SUSY,
respectively. In this case the internal symmetry group has the form 
$\mbox{SU}(2)_I$, where the indices are taken to be 
$ i,j \in \{1 , 2 \}$, and 
the supercharge satisfies a SU(2) Majorana condition given by
(\ref{eq:majorana}) 
and a two components spinor representation of the algebra (\ref{eq:SUSY al. with central charge}) is discussed in \ref{ch:algebra}. 
Throughout this paper we consider the Euclidean flat spacetime.  

We consider a twist of the algebra (\ref{eq:SUSY al. with central
charge}) by using the Dirac-K\"ahler twisting procedure.
This procedure means that a SO(4)$_I$ R-symmetry is identified with
SO(4) Euclidean rotation symmetry. 
Thus this procedure is not available for SU(2)$_I$ R-symmetry directly.
We consider a twist of $N$=4 SUSY algebra with SO(4)$_I$ R-symmetry. 
Since SO(4) group satisfies the following relation,
\begin{eqnarray}
SO(4)\simeq SU(2)\otimes SU(2),
\end{eqnarray}  
we can extract the corresponding SU(2)$_I$ R-symmetry.

We define $N$=4 supercharge as $\mathbf{Q}_{\alpha i}$ to distinguish  $N$=4 supercharge from $N$=2 supercharge.
It should be noted that index \{i\} takes $i \in  \{ 1 \dots 4 \}$,
because the supercharge $\mathbf{Q}_{\alpha i}$ has index
of the SO(4)$_I$ R-symmetry. We apply the Dirac-K\"ahler mechanism to
the $N$=4 supercharge:

%
\begin{eqnarray}
\mathbf{Q}_{\alpha i} &=& \frac{1}{\sqrt{2}}(s + \gamma^\mu s_\mu + \frac{1}{2}
\gamma^{\mu\nu}s_{\mu\nu} + \tilde{\gamma}^\mu \tilde{s}_\mu + \gamma_5 
\tilde{s} )_{\alpha i},
\label{eq:N=4 DKC - normal}
\end{eqnarray}
where the index $\alpha$ and $i$ are the Euclidean rotation SO(4) symmetry
 indices and\\ $ \{s ,s_\mu, s_{\mu\nu}, \tilde{s}_\mu, \tilde{s} \}$ stand for the twisted supercharges. 
We redefine the twisted supercharges as the following form to extract the $N=2$ sector,
\begin{eqnarray}
\left\{\begin{array}{lcl}
s^{\pm}\equiv \frac{1}{\sqrt{2}}(s\pm\tilde{s})\\
s^{\pm}_{\mu}\equiv \frac{1}{\sqrt{2}}(s_\mu \pm \tilde{s}_\mu)\\
s^{\pm}_{\mu\nu}\equiv \frac{1}{\sqrt{2}}(s_{\mu\nu}\mp\frac{1}{2}
\epsilon_{\mu\nu\rho\sigma}s^{\rho\sigma}).
\end{array}\right.
\end{eqnarray}
The equation (\ref{eq:N=4 DKC - normal}) is rewritten as the following, 
\begin{eqnarray}
\mathbf{Q}_{\alpha i} &=& (s^+ P_+ + s^+ _\mu \gamma^\mu P_+ + \frac{1}{4} s^+ _A \gamma^A P_+ 
\nonumber\\
&&+ s^- P_- + s^- _\mu \gamma^\mu P_- + \frac{1}{4} s^- _A \gamma^A P_- 
)_{\alpha i},
\label{eq:N=4 DKC}
\end{eqnarray}
where the capital $\{A\}$ means anti-symmetric tensor indices $\{ \mu\nu
\}$, projection matrices $(P_\pm)_{ij}$ of the internal symmetry are
defined by $(P_\pm)_{ij}= \frac{1}{2}(1\pm\gamma_5)_{ij}$ and the second
rank tensors satisfy (anti-)self-dual conditions: 
$s^{\pm}_{\mu\nu} = \mp\frac{1}{2}\epsilon_{\mu\nu\rho\sigma}
s^{\pm\rho\sigma}$.

As we are considering the $N$=2 case, 
we extract the corresponding supercharges from the equation 
(\ref{eq:N=4 DKC}). We exclude the supercharges $\{s^- ,s^- _\mu, s^-
_{\mu\nu}\}$, applying $P_+$ to the equation (\ref{eq:N=4 DKC}),
\begin{eqnarray}
(\mathbf{Q} P_+)_{\alpha i} = (s^+ P_+ + s^+ _\mu \gamma^\mu P_+ + \frac{1}{4} s^+ _A \gamma^A P_+ )_{\alpha i}.
\end{eqnarray}
This means the chiral projection with respect to the internal
 $SO(4)$ symmetry. It should be noted that
 the indices $ i,j \in \{1,2,3,4 \}$ but $i \in \{3,4\}$ components are zero 
because of the form of the matrix $ P_+ = \left(
\begin{smallmatrix}
\bf{1}_{2 \times 2} & 0 \\
0 & 0
\end{smallmatrix}
\right)$. 
We then identify the chiral supercharges as the $N$=2 sector of the
supercharges, 
\begin{eqnarray}
Q_{\alpha i} \equiv (\mathbf{Q}P_+)_{\alpha i}.
\end{eqnarray}
We define the conjugate supercharge as follows:
\begin{eqnarray}
\overline{Q}_{i \alpha} \equiv (P_+ C^{-1} \mathbf{Q}C)_{i \alpha},
\end{eqnarray}
where $C$ is the charge conjugation matrix and satisfies
\begin{align}
\gamma^T &=C\gamma_\mu C^{-1},& C^T=-C.
\end{align}
The twisted supercharges satisfy the following relations,
\begin{align}
s^+ &=\frac{1}{2} \mbox{Tr} (\mathbf{Q} P_+), & s^+ _\mu &= \frac{1}{2} \mbox{Tr}(\gamma_\mu
\mathbf{Q} P_+), & s^+ _A &=-\frac{1}{2} \mbox{Tr}(\gamma_A \mathbf{Q}
 P_+).
\label{eq:TSC}
\end{align}
The algebra (\ref{eq:SUSY al. with central charge}) is equivalently
rewritten as 
\begin{eqnarray}
\{(\mathbf{Q} P_+)_{\alpha i},(\mathbf{Q} P_+)_{\beta j} \}= 2(P_+ C P_+ )_{ji} (\gamma^\mu C^{-1} )_{\alpha\beta} +2(P_+)_{ij} \delta_{\alpha\beta}Z.
\end{eqnarray}
We calculate the anti-commutation relations of these twisted
supercharges,
\begin{align}
\{ s^+,s^+ _\mu\} &= P_\mu ,& \{ s^+ _A , s^+ _\mu \} &= -\delta^+ _{A,\mu\nu} P^\nu,& \{s^+,s^+_A\} &= 0, \nonumber \\
\{s^+,s^+\} &= Z ,& \{s^+_\mu,s^+_\nu\} &= \delta_{\mu\nu} Z ,& \{s^+_A,s^+_B\} &=  \delta^+ _{A,B} Z,
\label{eq:TSUSY al. with central charge}
\end{align}
where $\delta^+ _{\mu\nu, \rho\sigma}$ is defined as 
$\delta^+ _{\mu\nu, \rho\sigma}\equiv
\delta_{\mu\rho}\delta_{\nu\sigma} - \delta_{\mu\sigma}\delta_{\nu\rho}
- \epsilon_{\mu\nu\rho\sigma}$.
This algebra is $N$=2 twisted supersymmetry algebra with the central charge.


We construct a $N$=2 twisted superspace formalism based on the algebra
(\ref{eq:TSUSY al. with central charge}). Thus the $N$=2 twisted
superspace with the central charge consists of $\{x^\mu, \theta^+,
\theta^+_\mu, \theta^+ _{\mu\nu},z\} $ coordinates, where $x^\mu$, and  $z$ are
bosonic and  $ \theta^+, \theta^+_\mu $ and $ \theta^+ _{\mu\nu}$ are
fermionic.
We define differential operators corresponding with supercharges (\ref{eq:TSC}) as follows:
\begin{eqnarray}
\mathcal{Q}^+ &=& \frac{\partial}{\partial \theta^+} +\frac{i}{2} \theta^{+\mu}\partial_\mu +\frac{i}{2}\theta^+ \frac{\partial}{\partial z} \nonumber, \\
\mathcal{Q}^+ _{\mu} &=& \frac{\partial}{\partial \theta^{+\mu}} +\frac{i}{2} \theta^{+}\partial_\mu-\frac{i}{2}\theta^+ _{\mu\nu}\partial^\nu +\frac{i}{2}\theta^+_\mu \frac{\partial}{\partial z}, \nonumber \\
\mathcal{Q}^+ _A &=& \frac{\partial}{\partial \theta^{+A}} 
-\frac{i}{2} \delta^+ _{A,\rho\sigma}\theta^{+\rho}\partial^\sigma
 +\frac{i}{2}\theta^+_A \frac{\partial}{\partial z}, \nonumber \\
\mathcal{Z}&=& -i\frac{\partial}{\partial z},
\end{eqnarray}
where $z$ is a bosonic parameter corresponding to the central charge $Z$.
These differential operators satisfy the following relations, 
\begin{align}
\{ \mathcal{Q}^+,\mathcal{Q}^+ _\mu\} &= i\partial_\mu ,& 
\{ \mathcal{Q}^+ _A , \mathcal{Q}^+ _\mu \} &= -i\delta^+ _{A,\mu\nu} \partial^\nu,& \{\mathcal{Q}^+,\mathcal{Q}^+_A\}&=0,\nonumber \\
\{\mathcal{Q}^+,\mathcal{Q}^+\} &=- \mathcal{Z} ,& \{\mathcal{Q}^+_\mu,\mathcal{Q}^+_\nu\}&= -\delta_{\mu\nu} \mathcal{Z} ,& \{\mathcal{Q}^+_A,\mathcal{Q}^+_B\} &= - \delta^+ _{A,B} \mathcal{Z}. 
\end{align}
We introduce another set of differential operators $\{D^+_I\}$ which
anticommute with the differential operators $\{Q^+_I\}$,
\begin{eqnarray}
\mathcal{D}^+ &=& \frac{\partial}{\partial \theta^+} -\frac{i}{2} \theta^{+\mu}\partial_\mu -\frac{i}{2}\theta^+ \frac{\partial}{\partial z} \nonumber, \\
\mathcal{D}^+ _{\mu} &=& \frac{\partial}{\partial \theta^{+\mu}} -\frac{i}{2} \theta^{+}\partial_\mu  + \frac{i}{2}\theta^+ _{\mu\nu}\partial^\nu -\frac{i}{2}\theta^+_\mu \frac{\partial}{\partial z}, \nonumber \\
\mathcal{D}^+ _A &=& \frac{\partial}{\partial \theta^{+A}} 
+\frac{i}{2} \delta^+ _{A,\rho\sigma}\theta^{+\rho}\partial^\sigma
 -\frac{i}{2}\theta^+_A \frac{\partial}{\partial z}.
\end{eqnarray}

Next we consider the general characteristics of a superfield with the central
charge. If we know the TSUSY transformations of component fields,  we
define the general superfield $\Phi$ as follows,
\begin{eqnarray}
\Phi (x^\mu, \theta^I, z) = e^{\delta_{\theta^+,z}}\phi(x^\mu),
\label{eq:def.superfield}
\end{eqnarray}
where $\phi(x^\mu)$ is any field and
$\delta_{ \theta^+ ,z } = \theta^+ s^+ + \theta^{+\mu} s^+_\mu +
\frac{1}{4}\theta^{+A} s^+_A +iz Z=\delta_{ \theta^+ }+izZ $.
Since the operator $Z$ commutes all the operator $\{s^+_I\}$, the
superfield is given by
\begin{eqnarray}
\Phi(x^\mu,\theta^{+I},z) &=& e^{izZ}e^{\delta_{\theta^+}} \phi(x^\mu).
\end{eqnarray}
We can define the following operator, 
\begin{eqnarray}
\cosh(z\partial) &\equiv& 1+ \frac{1}{2}z^2 \partial^2 + \frac{1}{4!} z^4 \partial^4+ \frac{1}{6!} z^6 \partial^6 + \cdots ,\nonumber \\
\frac{\sinh(z\partial)}{\partial} &\equiv& z + \frac{1}{3!} z^3 \partial^2+ \frac{1}{5!} z^5 \partial^4 +  \frac{1}{7!}z^7 \partial^6 + \cdots.
\end{eqnarray}
Using these operators and the relation $Z^2 = -\partial^2$, $e^{izZ}$ is expressed as follows: 
\begin{eqnarray}
e^{izZ} = \cosh(z\partial) + i\frac{\sinh(z\partial)}{\partial}Z.
\end{eqnarray}
Thus the general superfield is given by
\begin{eqnarray}
\Phi(x^\mu,\theta^I,z) &=&  \cosh(z\partial)  e^{\delta_{\theta^+}} \phi(x^\mu) +  i\frac{\sinh(z\partial)}{\partial} Z  e^{\delta_{\theta^+}} \phi(x^\mu) \nonumber \\
&=&  \cosh(z\partial)  \Psi(x^\mu,\theta^I) + i\frac{\sinh(z\partial)}{\partial}\tilde{\Psi}(x^\mu,\theta^I),
\end{eqnarray}
where $\Psi(x^\mu , \theta^I)\equiv e^{\delta_{\theta^+}}
\phi(x^\mu),\quad \tilde{\Psi}(x^\mu ,\theta^I)\equiv Z
\Psi(x^\mu,\theta^I)$.
Since the general superfield is an infinite series with respect to $z$,
one may think that there are infinite number of component fields.
The superfield, however, has a finite number of component fields because
$\Psi$ and $\tilde{\Psi}$ depend only on $\theta^+$, $\theta^+_\mu$ and
$\theta^+_{\mu\nu}$. 

We consider a superfield which represents the hypermultiplet. 
Since the general superfield has many component fields as stated above,
we need to eliminate superfluous fields.
We then use the R-symmetry in order to impose a condition on the
superfield.   
We consider a superfield 
\footnote[1]{We can think of another construction of the
hypermultiplet. For example, we can introduce $\mathcal{V}$ and
$\mathcal{V_A}$ which transforms $ R^+_A \mathcal{V} =- \frac{i}{2}
\mathcal{V}_A$ and $ R^+_A  \mathcal{V}_B = \frac{i}{2} \delta^+ _{A,B}\
\mathcal{V} -\frac{i}{8} \Gamma^+ _{A,B,C}\ \mathcal{V}^{+C}  $,
respectively. We can further introduce a superfield with spinor index
\cite{AL}, but it is difficult to deal with this type of superfield in our
tensor formalism.}    
with a vector index ; $\mathcal{V}_\mu$. 
We also introduce a R-transformations for the superfield $\mathcal{V}_\mu$
and supercharges:  
\begin{eqnarray}
R^+ _A \mathcal{V}_\mu &=& -\frac{i}{2}\delta^+ _{A,\mu\nu}\mathcal{V}^\nu \nonumber ,\\
R^+_A s^+ &=&- \frac{i}{2} s^+_A ,\nonumber \\
R^+_A  s^+_B &=& \frac{i}{2} \delta^+ _{A,B}\ s^+ -\frac{i}{8} \Gamma^+ _{ABC}\ s^{+C} ,\nonumber \\
R^+_A s^+_\mu &=& -\frac{i}{2}\delta^+ _{A,\mu\nu} s^{+\nu} ,
\label{eq:R-symmetry}
\end{eqnarray}
where $\Gamma^+_{ABC}$ is an anti-symmetric tensor defined in the
Appendix. The $\{\mathcal{D}\}$ operators transform in the same
manner with respect to the supercharges. 
We can find the following R-invariant terms,
\begin{align}
R^+_A( \mathcal{D} ^+ \mathcal{V}_\mu +\mathcal{D}^+_{\mu\nu}
 \mathcal{V}^\nu) &=0, & R^+_A ( \mathcal{D} ^+_\mu  \mathcal{V}^\mu) &=
 0,& R^+_A( \delta^- _{\mu\nu,\rho\sigma}\mathcal{D}^{+\rho}
 \mathcal{V}^\sigma) &=0. 
\label{eq:R-inv}
\end{align}
We impose the following condition,
\begin{eqnarray}
R^+_A( \mathcal{D}^+ _\mu \mathcal{V}_\nu )=0.
\label{eq:con}
\end{eqnarray}
We expand  $R^+_A( \mathcal{D}^+ _\mu \mathcal{V}_\nu ) $ as follows:
\begin{eqnarray}
R^+_A( \mathcal{D}^+_\mu \mathcal{V}_\nu) &=&R^+_A \Big{(} \frac{1}{4}\delta_{\mu\nu}\mathcal{D}^{+\rho} 
\mathcal{V}_\rho  +\frac{1}{2}( \mathcal{D}^+_\mu \mathcal{V}_\nu + \mathcal{D}^+_\nu \mathcal{V}_\mu  -\frac{1}{2}\delta_{\mu\nu}\mathcal{D}^{+\rho} \mathcal{V}_\rho ) \nonumber  \\
& &\hspace{9mm}+\frac{1}{4}\delta^+ _{\mu\nu,\rho\sigma} \mathcal{D}^{+\rho} \mathcal{V}^\sigma +\frac{1}{4}\delta^- _{\mu\nu,\rho\sigma} \mathcal{D}^{+\rho} \mathcal{V}^\sigma \Big{)} \nonumber \\
&=&R^+_A \Big{(}   \frac{1}{2}( \mathcal{D}^+_\mu \mathcal{V}_\nu + \mathcal{D}^+_\nu \mathcal{V}_\mu  -\frac{1}{2}\delta_{\mu\nu}\mathcal{D}^{+\rho} \mathcal{V}_\rho )
+\frac{1}{4}\delta^+ _{\mu\nu,\rho\sigma} \mathcal{D}^{+\rho} \mathcal{V}^\sigma  \Big{)},
\label{eq:derive-const}
\end{eqnarray}
where we use the equation (\ref{eq:R-inv}). The right hand side of
equation (\ref{eq:derive-const}) must be zero because of the constraint
(\ref{eq:con}).
 We can similarly find a constraint by using the first equation of
 (\ref{eq:R-inv}). 
The superfield satisfies the following constraints,
\begin{eqnarray}
& &\mathcal{D}^+_A \mathcal{V}_\mu + \delta^+ _{A,\mu\nu} \mathcal{D}^+ \mathcal{V}^\nu = 0, \nonumber  \\
& & \mathcal{D}^+_\mu \mathcal{V}_\nu + \mathcal{D}^+_\nu \mathcal{V}_\mu = \frac{1}{2} \delta_{\mu\nu} \mathcal{D}^{+\rho} \mathcal{V}_\rho,  \nonumber \\
& &\delta^+ _{A,\mu\nu} \mathcal{D}^{+\mu}  \mathcal{V}^\nu  = 0.
\label{eq:constraint}
\end{eqnarray}
We then derive the following relations from the constraints
(\ref{eq:constraint}),
\begin{eqnarray}
\mathcal{D}^+ \mathcal{D}^{+\mu} \mathcal{V}_\mu &=& -2i \partial^\mu \mathcal{V}_\mu, \\
\delta^- _{A,\mu\nu} \mathcal{D}^+ \mathcal{D}^{+\mu} \mathcal{V}^\nu  &=& -2i\delta^- _{A,\mu\nu} \partial^\mu \mathcal{V}^\nu, \\
\mathcal{D}^+_\mu  \mathcal{D}^{+\nu} \mathcal{V}_\nu &=& 2\mathcal{Z} \mathcal{V}_\mu , \\
\delta^- _{A,\rho\sigma} \mathcal{D}^+_\mu \mathcal{D}^{+\rho} \mathcal{V}^\sigma &=& 2\delta^- _{A,\mu\nu} \mathcal{Z}\mathcal{V}^\nu , \\
\mathcal{Z} \mathcal{D}^+_\mu \mathcal{V}_\nu &=& -i \delta_{\mu\nu} \partial^\rho \mathcal{D}^+ \mathcal{V}_\rho  -i\delta^- _{\mu\nu,\rho\sigma} \partial^\rho \mathcal{D}^+ \mathcal{V}^\sigma , \\
\mathcal{Z}\mathcal{D}^+ \mathcal{V}_\mu &=& -\frac{i}{4}\partial_\mu \mathcal{D}^{+\nu} \mathcal{V}_\nu  +\frac{i}{4}\delta^- _{\mu\nu,\rho\sigma} \partial^\nu \mathcal{D}^{+\rho} \mathcal{V}^\sigma,  \\
\mathcal{Z}^2\mathcal{V}_\mu &=&-\partial^2 \mathcal{V}_\mu. 
\end{eqnarray}
It is difficult to generally solve the constraints (\ref{eq:constraint}).
We find transformation laws of TSUSY with respect to component fields by
the following method. We check whether these transformations satisfy the
algebras (\ref{eq:TSUSY al. with central charge}).

We define component fields of the hypermultiplet as:
\begin{align}
\mathcal{V}_\mu| &= V_\mu, &  \mathcal{D}^+ \mathcal{V}_\mu | &=\tilde{\psi}_\mu,&  
\frac{1}{4}\mathcal{D}^{+\mu} \mathcal{V}_\mu | &= \tilde{\eta},\nonumber   \\
\frac{1}{4}\delta^- _{A,\mu\nu} \mathcal{D}^{+\mu} \mathcal{V}^\nu| &= -\chi^- _A , & Z \mathcal{V}_\mu | &= K_\mu , 
\end{align}
where $|$ means the lowest component of $\theta$'s.
One can derive transformation laws of the hypermultiplet as follows:
\begin{eqnarray}
s^+ V_\mu = \mathcal{Q}^+ \mathcal{V}_\mu | =\mathcal{D}^+ \mathcal{V}_\mu |= \tilde{\psi}_\mu, 
\end{eqnarray}
where the first equality is a definition of the SUSY transformation
, we use the fact that the operator $\mathcal{Q}^+$ is equivalent to
$\mathcal{D}^+ $ at the lowest component of $\theta$'s  at the second
one and we use the definition of the field $\tilde{\psi}_\mu$ at the
last one. 
\begin{eqnarray}
s^+_\mu V_\nu &=& \mathcal{Q}^+_\mu \mathcal{V}_\nu |  = \mathcal{D}^+_\mu \mathcal{V}_\nu |  \nonumber \\
&=& \frac{1}{2}(\mathcal{D}^+_\mu\mathcal{V}_\nu+ \mathcal{D}^+_\nu \mathcal{V}_\mu)| + \frac{1}{2}( \mathcal{D}^+_\mu\mathcal{V}_\nu-  \mathcal{D}^+_\nu \mathcal{V}_\mu)| \nonumber \\
&=& \frac{1}{4}\delta_{\mu\nu} \mathcal{D}^{+\rho} \mathcal{V}_\rho | 
    +\frac{1}{4}\delta^+ _{\mu\nu,\rho\sigma} \mathcal{D}^{+\rho} \mathcal{V}^\sigma | +\frac{1}{4}\delta^- _{\mu\nu,\rho\sigma} \mathcal{D}^{+\rho} \mathcal{V}^\sigma | \nonumber \\
&=& \delta_{\mu\nu} \tilde{\eta} -\chi^- _{\mu\nu},
\end{eqnarray}
where we use constraints (\ref{eq:constraint}) and definitions of
$\tilde{\eta}$ and $\chi^-_{\mu\nu}$. 
We find the other transformations of component fields. 
We show the $N$=2 TSUSY transformations of the hypermultiplet
in Table \ref{tb:trhy1}. 
\begin{table}[htbp]
\[\hspace{-15mm}
\begin{array}{|c||c|c|c|c|}
\hline
 & s^+ & s^+_\mu & s^+_A & Z \\
\hline
V_\nu & \tilde{\psi}_\nu & \delta_{\mu\nu} \tilde{\eta}-\chi^- _{\mu\nu} &
-\delta^+ _{A,\nu\rho}\tilde{\psi}^\rho &K_\nu \\
\tilde{\psi}_\nu & \frac{1}{2}K_\nu &\frac{i}{2} (\delta^- _{\mu\nu,\rho\sigma}\partial^\rho V^\sigma + \delta_{\mu\nu} \partial^\rho V_{\rho} -2\partial_{\mu} V_{\nu}) &\frac{1}{2}\delta^+ _{A,\nu\rho}K^\rho &-i(\partial_\nu \tilde{\eta} + \partial^\rho \chi^- _{\nu\rho}) \\
\tilde{\eta} & -\frac{i}{2} \partial^\mu V_\mu & \frac{1}{2} K_\mu&-\frac{i}{2}\delta^+ _{A,\rho\sigma}\partial^\rho V^\sigma &- i \partial^\rho \tilde{\psi}_\rho \\ 
\chi^- _B &\frac{i}{2}\delta^-_{B,\rho\sigma}\partial^\rho V^{\sigma} &-\frac{1}{2} \delta^-_{B,\mu\rho}K^\rho  &-\frac{i}{2}\delta^+ _{A,\rho\nu}\delta^- _{B,\sigma}{}^\nu \partial^\rho V^\sigma  &i\delta^- _{B,\rho\sigma} \partial^\rho \tilde{\psi}^\sigma \\
K_\nu &-i(\partial_\nu \tilde{\eta} +\partial^\rho \chi^- _{\nu\rho}) &-i(\delta^- _{\mu\nu,\rho\sigma} \partial^\rho \tilde{\psi}^\sigma +\delta_{\mu\nu} \partial^\rho \tilde{\psi}_\rho)& i \delta^+ _{A,\nu\rho}(\partial^\rho \tilde{\eta} +\partial^\sigma \chi^{- \rho} {}_\sigma) &-\partial^2V_\nu \\
\hline
\end{array}
\]
\caption{$N$=2 TSUSY transformations of the Hypermultiplet.}
\label{tb:trhy1}
\end{table}
Since these transformations satisfy the algebra
(\ref{eq:SUSY al. with central charge}), this multiplet is one of the
solutions of the constraint (\ref{eq:constraint}). We can construct
the superfield $\mathcal{V}_\mu$ by applying the equation
(\ref{eq:def.superfield}).   

We consider a TSUSY invariant action. 
Taking the highest component of $\theta$'s out of a superfield does not
lead to a TSUSY invariant action.
The reason is that the derivative with respect to $z$ is added to these
supercharges $\{ \mathcal{Q} \}$. 
For example, $\mathcal{Q}^+$ is the following form,
\begin{eqnarray}
\mathcal{Q}^+ = \frac{\partial}{\partial \theta^+} +\frac{i}{2}\theta^{+\mu} \partial_\mu +\frac{i}{2} {\theta}^+ \frac{\partial}{\partial z}.
\end{eqnarray}
The first term does not contribute to the highest component of a
superfield, the second one is a total derivative term and
the last one is not a total derivative term. We recognize that the superfield
to which we apply $\mathcal{Q}^+$ is non-total derivative term.
The other charges have the same
structure. Thus the highest component of the superfield is not TSUSY
invariant. 
We find a TSUSY invariant action to take the lowest component of
$\theta$'s. 
The explicit form of the action is the following form, 
\begin{eqnarray}
S&=&\frac{1}{2}\int d^4x \Big{(} -\frac{1}{6}\delta^+ _{\mu\nu,\rho\sigma} \mathcal{D}^{+\mu} \mathcal{D}^{+\nu} (\mathcal{V}^\rho \mathcal{Z}\mathcal{V}^\sigma ) -\frac{1}{12}\delta^+ _{\mu\nu,\rho\sigma}
 \mathcal{D}^{+\mu \alpha} \mathcal{D}^{+\nu} {} _{\alpha} (\mathcal{V}^\rho \mathcal{Z}\mathcal{V}^\sigma ) \Big{)} | \nonumber \\
&=& \int d^4 x \Big{(} V^\mu \partial^2 V_\mu +4i\tilde{\psi}^\mu (\partial_\mu \tilde{\eta}+\partial^\nu \chi^- _{\mu\nu} ) +K^\mu K_\mu \Big{)}.
\end{eqnarray}
This action possesses the TSUSY invariance of Table \ref{tb:trhy1} and
the following R-symmetry, 
\begin{align}
R^+_A V_\mu &= -\frac{i}{2}\delta^+ _{A,\mu\nu}V^\nu, & R^+_A K_\mu &=
 -\frac{i}{2}\delta^+ _{A,\mu\nu}K^\nu, & R^+_A\tilde{\psi} &=R^+_A
 \tilde{\eta}=  R^+_A\chi^- _B =0 .
\end{align}

\setcounter{equation}{0}
\renewcommand{\theequation}{\arabic {section}.\arabic{equation}}
\section{Hypermultiplet coupling to Donaldson-Witten theory}
We will construct the Donaldson-Witten theory coupled to the
hypermultiplet. The constraints (\ref{eq:constraint}) must be
covariantized because of the coupling to the gauge field.
\begin{eqnarray}
& &\nabla^+_A \mathcal{V}_\mu + \delta^+ _{A,\mu\nu} \nabla^+ \mathcal{V}^\nu = 0, \nonumber  \\
& & \nabla^+_\mu \mathcal{V}_\nu + \nabla^+_\nu \mathcal{V}_\mu = \frac{1}{2} \delta_{\mu\nu} \nabla^{+\rho} \mathcal{V}_\rho,  \nonumber \\
& &\delta^+ _{A,\mu\nu} \nabla^{+\mu}  \mathcal{V}^{+\nu}  = 0,
\label{eq:covariant-constraint}
\end{eqnarray}
where $ \nabla^+ \equiv \mathcal{D}^+ -i\Gamma$, $\nabla^+_\mu \equiv
\mathcal{D}^+_\mu -i\Gamma_\mu$, $\nabla^+_A \equiv
\mathcal{D}^+_A -i\Gamma_A$ and  $\{\Gamma, \Gamma_{\mu}, \Gamma_{A} \}$ 
are connection superfields. We derive the following relations from the
constraint (\ref{eq:covariant-constraint}) and the Jacobi identities, 
\begin{eqnarray}
\nabla^+ \nabla^{+\mu} \mathcal{V}_\mu &=& -2i \nabla^{\underline{\mu}} \mathcal{V}_\mu, \\
\delta^- _{A,\mu\nu} \nabla^+ \nabla^{+\mu} \mathcal{V}^\nu  &=& -2i\delta^- _{A,\mu\nu} \nabla^{\underline{\mu}} \mathcal{V}^\nu, \\
\nabla^+_\mu  \nabla^{+\nu} \mathcal{V}_\nu &=& 2\mathcal{Z} \mathcal{V}_\mu -2i[\mathcal{W},\mathcal{V}_\mu], \\
\delta^- _{A,\rho\sigma} \nabla^+_\mu \nabla^{+\rho} \mathcal{V}^\sigma &=& 2\delta^- _{A,\mu\nu} \mathcal{Z}\mathcal{V}^\nu -2i\delta^- _{A,\mu\nu} [\mathcal{W},\mathcal{V}^\nu], \\
\mathcal{Z} \nabla^+_\mu \mathcal{V}_\nu &=& -i \delta_{\mu\nu} \nabla^{\underline{\rho}} \nabla^+ \mathcal{V}_\rho  -i\delta^- _{\mu\nu,\rho\sigma} \nabla^{\underline{\rho}} \nabla^+ \mathcal{V}^\sigma +i[\mathcal{F},\nabla^+_{\mu}\mathcal{V}_\nu] \nonumber \\
& &+\frac{i}{2}\delta_{\mu\nu}[\nabla^{+\rho} \mathcal{F},\mathcal{V}_\rho]
+\frac{i}{2}\delta^- _{\mu\nu,\rho\sigma}[\nabla^{+\rho} \mathcal{F} , \mathcal{V}^\sigma], \\
\mathcal{Z}\nabla^+ \mathcal{V}_\mu &=& -\frac{i}{4}\nabla_{\underline{\mu}} \nabla^{+\nu} \mathcal{V}_\nu  +\frac{i}{4}\delta^- _{\mu\nu,\rho\sigma} \nabla^{\underline{\nu}} \nabla^{+\rho} \mathcal{V}^\sigma+\frac{i}{2}[\nabla^+ \mathcal{W} ,\mathcal{V}_\mu]+\frac{i}{2}[\nabla^+_{\mu\rho} \mathcal{W} ,\mathcal{V}^\rho]\nonumber \\
& &+i[\mathcal{W},\nabla^+ \mathcal{V}_\mu] , \\ 
\mathcal{Z}^2\mathcal{V}_\mu &=&-\nabla^{\underline{\nu}}\nabla_{\underline{\nu}} \mathcal{V}_\mu
+\frac{i}{4}\{\nabla^+_\mu \mathcal{F},\nabla^{+\nu} \mathcal{V}_\nu \}-\frac{i}{4}\delta^-_{\mu\nu.\rho\sigma}\{\nabla^{+\nu} \mathcal{F} , \nabla^{+\rho} \mathcal{V}^\sigma\} \nonumber \\& & + i\{\nabla^+ \mathcal{W} ,\nabla^+ \mathcal{V}_\mu\}-i\{\nabla^+_{\mu\nu}\mathcal{W} ,\nabla^+ \mathcal{V}^\nu\}
+\frac{i}{4}\delta^+ _{\mu\nu,\rho\sigma}[\nabla^{+\rho} \nabla^{+\sigma} \mathcal{F},  \mathcal{V}^\nu] \nonumber \\
 & &+ i[\mathcal{W},\mathcal{Z} \mathcal{V}_\mu] +i[\mathcal{F},\mathcal{Z} \mathcal{V}_\mu] +\frac{1}{2}[\mathcal{F},[\mathcal{W}, \mathcal{V}_\mu]]+\frac{1}{2}[\mathcal{W},[\mathcal{F}, \mathcal{V}_\mu]], 
\end{eqnarray}
where $\mathcal{F}$ and $\mathcal{W}$ are bosonic curvature superfields
with respect to the vector multiplet \cite{KKM} and we use the following modified
commutation relations,
\begin{align}
\{\nabla^+,\nabla^+\} &= -i\mathcal{F} + \mathcal{Z} ,& \{\nabla^+_A,\nabla^+_B\} &= -i \delta^+ _{A,B}\mathcal{F}+\delta^+ _{A,B}\mathcal{Z} , \nonumber \\
\{\nabla^+_\mu,\nabla^+_\nu\} &= -i\delta_{\mu\nu}\mathcal{W}+ \delta_{\mu\nu}\mathcal{Z} ,&\{\nabla^+,\nabla^+_\mu\} &= -i\nabla_{\underline{\mu}} , \nonumber \\
\{\nabla^+_A,\nabla^+_\mu\} &= i\delta^+ _{A,\mu\nu}\nabla^{\underline{\nu}} ,&
{[}\nabla^+, \nabla_{\underline{\mu}}{]} &= -\frac{i}{2}\nabla^+_\mu \mathcal{F}, \nonumber \\
{[}\nabla^+_A, \nabla_{\underline{\mu}}{]} &= -\frac{i}{2}\delta^+ _{A,\mu\nu}\nabla^{+\nu} \mathcal{F},&
{[}\nabla^+_\mu, \nabla_{\underline{\nu}}{]} &= -\frac{i}{2} \delta_{\mu\nu}\nabla^+  \mathcal{W} +\frac{i}{2}\nabla^+_{\mu\nu}\mathcal{W}.
\end{align}
The curvature superfields $\mathcal{F}$ and $\mathcal{W}$ commute with the central charge $\mathcal{Z}$.
We define component fields of superfields $\mathcal{F}$, $\mathcal{W}$
and $\mathcal{V}_\mu$ as follows:
\begin{align}
\mathcal{F}| &= \phi, & \nabla_\mu \mathcal{F}| &= C_\mu, &
 \frac{1}{4}\delta^+ _{A,\mu\nu} \nabla^\mu\nabla^\nu
\mathcal{F}| &=- \phi^+_{A}, \nonumber \\
\mathcal{W}| &= \overline{\phi}, & \nabla_A \mathcal{W}| &=\chi^+ _A, & 
\nabla \mathcal{W}| &=\chi, \nonumber \\
\mathcal{V}_\mu| &= V_\mu, &  \nabla^+ \mathcal{V}_\mu | &=\tilde{\psi}_\mu,&  
\frac{1}{4}\nabla^{+\mu} \mathcal{V}_\mu | &= \tilde{\eta},\nonumber   \\
\frac{1}{4}\delta^- _{A,\mu\nu} \nabla^{+\mu} \mathcal{V}^\nu| &= -\chi^- _A
 , & \mathcal{Z} \mathcal{V}_\mu | &= K_\mu .  
\end{align}
One can similarly derive transformation laws of the hypermultiplet as
follows: 
\begin{eqnarray}
s^+ V_\mu &=& \nabla^+ \mathcal{V}_\mu | = \tilde{\psi}_\mu, \\
s^+_\mu V_\nu &=& \nabla^+_\mu \mathcal{V}_\nu | = \frac{1}{2}(\nabla^+_\mu\mathcal{V}_\nu+\nabla^+_\nu \mathcal{V}_\mu)| + \frac{1}{2}(\nabla^+_\mu\mathcal{V}_\nu- \nabla^+_\nu \mathcal{V}_\mu)| \nonumber \\
&=& \frac{1}{4}\delta_{\mu\nu}\nabla^{+\rho} \mathcal{V}_\rho | 
    +\frac{1}{4}\delta^+ _{\mu\nu,\rho\sigma} \nabla^{+\rho} \mathcal{V}^\sigma | +\frac{1}{4}\delta^- _{\mu\nu,\rho\sigma} \nabla^{+\rho} \mathcal{V}^\sigma | \nonumber \\
&=& \delta_{\mu\nu} \tilde{\eta} -\chi^- _{\mu\nu},
\end{eqnarray}
where we use the constraints (\ref{eq:covariant-constraint}) and the
definition of $\tilde{\eta}$ and $\chi^-_{\mu\nu}$. It should be noted
that the lowest components of the fermionic superconnections vanish because
we impose the Wess-Zumino gauge.
We show the $N$=2 TSUSY transformations of the hypermultiplets
in Table \ref{tb:trhy2}. The commutation relations of the twisted
supercharges are given in the following form,
\begin{align}
\{s^+,s^+\}\varphi &= Z\varphi -i[\phi,\varphi],& \{s^+,s^+_\mu\}\varphi &= -iD_\mu \varphi,& {[}s^+,Z{]} \varphi &= 0, \nonumber
 \\
\{s^+_\mu,s^+_\nu\}\varphi &=\delta_{\mu\nu} Z\varphi -i\delta_{\mu\nu} [\overline{\phi},\varphi],& \{s^+_\mu,s^+_A\}\varphi &=
 i\delta^+ _{\mu\nu,A} D^\nu\varphi,&
 {[}s^+_\mu,Z{]} \varphi &= 0, \nonumber \\
\{s^+_A,s^+_B\} \varphi &= \delta^+ _{A,B} Z \varphi -i\delta^+
 _{A,B}[\phi,\varphi],& \{s^+,s^+_A\}\varphi &= 0,
& {[}s^+_A,Z{]} \varphi &= 0,
\end{align}
where $\varphi = V_\mu,\tilde{\psi}_\mu , \tilde{\eta}, \chi^- _A ,
K_\mu$ and $D_{\mu}\varphi =\partial_\mu \varphi -i[\omega_\mu,
\varphi]$. Thus these algebras are closed at the off-shell level up to
the gauge transformation.
\begin{table}[htbp]
\[
\begin{array}{|c||c|c|}
\hline
 & s^+ & s^+_\mu  \\
\hline
V_\nu & \tilde{\psi}_\nu & \delta_{\mu\nu} \tilde{\eta}-\chi^- _{\mu\nu}  \\
\tilde{\psi}_\nu & \frac{1}{2}K_\nu -\frac{i}{2}[\phi,V_\nu] &\frac{i}{2} (\delta^- _{\mu\nu,\rho\sigma}D^\rho V^\sigma + \delta_{\mu\nu} D^\rho V_{\rho} -2D_{\mu} V_{\nu}) \\
\tilde{\eta} & -\frac{i}{2} D^\mu V_\mu & \frac{1}{2} K_\mu -\frac{i}{2}[\overline{\phi},V_\mu]\\  
\chi^- _B &\frac{i}{2}\delta^-_{B,\rho\sigma}D^\rho V^{\sigma} &-\frac{1}{2} \delta^-_{B,\mu\rho}K^\rho +\frac{i}{2}\delta^-_{B,\mu\nu}[\overline{\phi},V^\nu] \\
K_\nu &-i(D_\nu \tilde{\eta} +D^\rho \chi^- _{\nu\rho})+\frac{i}{2}[\chi,V_\nu] &-i(\delta^- _{\mu\nu,\rho\sigma} D^\rho \tilde{\psi}^\sigma +\delta_{\mu\nu} D^\rho \tilde{\psi}_\rho)+i[\phi,\delta_{\mu\nu}\tilde{\eta}-\chi^- _{\mu\nu}]\\
 & +\frac{i}{2}[\chi^+ _{\nu\mu},V^\mu]+i[\overline{\phi},\tilde{\psi}_\nu] & +\frac{i}{2}\delta_{\mu\nu}[C^\rho,V_\rho] +\frac{i}{2}\delta^- _{\mu\nu,\rho\sigma}[C^\rho,V^\sigma]\\
\hline
\end{array}
\]
\vspace{5mm}
\[\hspace{-8mm}
\begin{array}{|c||c|c|}
\hline
 & s^+_A & Z \\
\hline
V_\nu  &
-\delta^+ _{A,\nu\rho}\tilde{\psi}^\rho &K_\nu \\
\tilde{\psi}_\nu &   \frac{1}{2}\delta^+ _{A,\nu\rho}K^\rho -\frac{i}{2}\delta^+ _{A,\nu\rho}[\phi,V^\rho]& -i(D_\nu \tilde{\eta} + D^\rho \chi^- _{\nu\rho})+\frac{i}{2}[\chi,V_\nu]+\frac{i}{2}[\chi^+_{\nu\mu},V^\mu]+i[\overline{\phi},\tilde{\psi}_\nu] \\
\tilde{\eta} &-\frac{i}{2}\delta^+ _{A,\rho\sigma}D^\rho V^\sigma &- i D^\rho \tilde{\psi}_\rho +\frac{i}{2}[C^\mu,V_\mu]+i[\phi,\tilde{\eta}]\\ 
\chi^- _B & -\frac{i}{2}\delta^+ _{A,\rho\nu}\delta^- _{B,\sigma}{}^\nu D^\rho V^\sigma  &i\delta^- _{B,\rho\sigma} D^\rho \tilde{\psi}^\sigma +i[\phi,\chi^- _B]-\frac{i}{2}\delta^-_{B,\mu\nu}[C^\mu,V^\nu]\\
K_\nu & i \delta^+ _{A,\nu\rho}(D^\rho \tilde{\eta} +D^\sigma \chi^{- \rho} {}_\sigma) -\frac{i}{2}\delta^+ _{A,\nu\mu}[\chi,V^\mu]  
&-D^\mu D_\mu V_\nu +i\{C^\mu, \delta_{\mu\nu}\tilde{\eta}+\chi^-_{\nu\mu}\} +i\{\delta_{\mu\nu}\chi-\chi^+ _{\nu\mu} ,\tilde{\psi}^\mu \}\\
&-\frac{i}{2}\delta^+ _{A,\nu\mu}[\chi^+ {}^\mu {}_{\rho},V^\rho]-i\delta^+ _{A,\nu\mu}[\overline{\phi},\tilde{\psi}^\mu] 
&-i[\phi^+ _{\nu\mu},V^\mu]+i[\phi+\overline{\phi},K_\nu] +\frac{1}{2}[\phi,[\overline{\phi},V_\nu]] +\frac{1}{2}[\overline{\phi},[\phi,V_\nu]]\\
\hline
\end{array}
\]
\caption{TSUSY transformation of the hypermultiplet coupling to
 the vector multiplet.}
\label{tb:trhy2}
\end{table}

A covariantized action has the following form,
\begin{eqnarray}
S_H&=&\frac{1}{4}\int d^4x  \mbox{Tr} \Big{(} -\frac{1}{6}\delta^+ _{\mu\nu,\rho\sigma} \nabla^{+\mu} \nabla^{+\nu} (\mathcal{V}^\rho \mathcal{Z}\mathcal{V}^\sigma ) -\frac{1}{12}\delta^+ _{\mu\nu,\rho\sigma}
 \nabla^{+\mu \alpha} \nabla^{+\nu} {} _{\alpha} (\mathcal{V}^\rho \mathcal{Z}\mathcal{V}^\sigma )   \Big{)} | \nonumber \\
&=& \int d^4 x\mbox{Tr} \Big{(} \frac{1}{2}V^\mu D^\nu D_\nu V_\mu +2i\tilde{\psi}^\mu (D_\mu \tilde{\eta}+D^\nu \chi^- _{\mu\nu} ) + \frac{1}{2}K^\mu K_\mu \nonumber \\
& &\hspace{20mm}+i \overline{\phi} \{\tilde{\psi}_\mu ,\tilde{\psi}^\mu\} +i\phi \{\tilde{\eta}, \tilde{\eta} \} +\frac{i}{4}\phi\{\chi^- _A ,\chi^{-A} \} \nonumber \\
& &\hspace{20mm}-iV^\mu\{C^\nu, \delta_{\mu\nu}\tilde{\eta} +\chi^- _{\mu\nu}\} -iV^\mu\{\delta_{\mu\nu} \chi -\chi^+ _{\mu\nu},\tilde{\psi}^\nu\} \nonumber \\
& &\hspace{20mm} -\frac{i}{2}\phi^+ _{\mu\nu}[V^\mu,V^\nu]
-\frac{1}{4}V^\mu[\phi,[\overline{\phi},V_\mu]]-\frac{1}{4}V^\mu[\overline{\phi},[\phi,V_\mu]]
\Big{)} .
\end{eqnarray}

The Donaldson-Witten theory is given by the following form:
\begin{eqnarray}
S_{DW} &=& \frac{1}{2} \int d^4x d^4\theta\ \mbox{Tr} \mathcal{F}^2 \nonumber \\
&=&\frac{1}{2} \int d^4 x\text{Tr}\Big{(}-\phi D^\mu D_\mu \overline{\phi} -iC^\mu(D_\mu \chi -D^\nu \chi^+ _{\mu\nu} )  + (F^- _{\mu\nu})^2 \nonumber \\
& &\hspace{23mm}+\frac{i}{2}\phi \{\chi,\chi\}+\frac{i}{8}\phi \{\chi^{+A},\chi^+ _A\} + \frac{i}{2} \overline{\phi} \{C^\mu,C_\mu\}\nonumber  \\
& &\hspace{23mm} +\frac{1}{4}[\phi,\overline{\phi}]^2 -\frac{1}{4}(\phi^+ _{\mu\nu} )^2 \Big{)}.
\label{eq:DW}
\end{eqnarray}
The superspace description of this action have been studied in \cite{AL}, \cite{KKM}.

We then obtain a new action, 
\begin{eqnarray}
S_{total} &=& S_H + S_{DW}\nonumber  \\
 &=&\int d^4 x\text{Tr}\Big{(}-\frac{1}{2}\phi D^\mu D_\mu \overline{\phi} + \frac{1}{2}V^\mu D^\nu D_\nu V_\mu +\frac{1}{2} (F^- _{\mu\nu})^2  +\frac{1}{8}[\phi,\overline{\phi}]^2 \nonumber \\
& &\hspace{0mm}+2i\tilde{\psi}^\mu (D_\mu \tilde{\eta}+D^\nu \chi^- _{\mu\nu} )+i\phi \{\tilde{\eta}, \tilde{\eta} \} +\frac{i}{4}\phi\{\chi^- _A ,\chi^{-A} \}+i \overline{\phi} \{\tilde{\psi}_\mu ,\tilde{\psi}^\mu\}  \nonumber \\
& &\hspace{0mm}-\frac{i}{2} C^\mu(D_\mu \chi -D^\nu \chi^+ _{\mu\nu} )+\frac{i}{4}\phi \{\chi,\chi\}+\frac{i}{16}\phi \{\chi^{+A},\chi^+ _A\} + \frac{i}{4} \overline{\phi} \{C^\mu,C_\mu\}\nonumber  \\
& &\hspace{0mm} 
 -iV^\mu\{C^\nu, \delta_{\mu\nu}\tilde{\eta} +\chi^- _{\mu\nu}\} -iV^\mu\{\delta_{\mu\nu} \chi -\chi^+ _{\mu\nu},\tilde{\psi}^\nu\} -\frac{1}{4}V^\mu[\phi,[\overline{\phi},V_\mu]]-\frac{1}{4}V^\mu[\overline{\phi},[\phi,V_\mu]]  \nonumber \\
& &\hspace{0mm}+ \frac{1}{2}K^\mu K_\mu  -\frac{1}{8}(\phi^+ _{\mu\nu} )^2  -\frac{i}{2}\phi^+ _{\mu\nu}[V^\mu,V^\nu]\Big{)}. 
\label{eq:total action}
\end{eqnarray}

In the ordinary supersymmetric theory $N$=4 vector multiplet consists
of $N$=2 vector multiplet and  hypermultiplet at on-shell level. 
By integrating out the auxiliary fields; $\phi^+ _A $ and $K_\mu$, we
can construct an on-shell invariant action. From the equation of motion auxiliary
fields are given by,
\begin{align}
\phi^+ _A &= - \frac{i}{2}\delta^+ _{A,\rho\sigma} [V^\rho,V^\sigma], &
 K_\mu&=0.
\end{align}
We redefine the field as
\begin{align}
C & \to 2C,& \chi & \to 2\chi, & \chi^+ _A &\to 2 \chi^+ _A, & \phi &\to
 i \phi , & \overline{\phi} \to  i \overline{\phi},
\end{align}
and obtain
\begin{eqnarray}
S_{total} &=& S_H + S_{DW}\nonumber  \\
 &=&\int d^4 x\text{Tr}\Big{(}\frac{1}{2}\phi D^\mu D_\mu \overline{\phi} + \frac{1}{2}V^\mu D^\nu D_\nu V_\mu +\frac{1}{2} (F^- _{\mu\nu})^2  +\frac{1}{8}[\phi,\overline{\phi}]^2 -\frac{1}{4}[V_\mu, V_\nu]^2\nonumber \\
& &\hspace{0mm}+2i\tilde{\psi}^\mu (D_\mu \tilde{\eta}+D^\nu \chi^- _{\mu\nu} ) -\phi \{\tilde{\eta}, \tilde{\eta} \} -\frac{1}{4}\phi\{\chi^- _A ,\chi^{-A} \}- \overline{\phi} \{\tilde{\psi}_\mu ,\tilde{\psi}^\mu\}  \nonumber \\
& &\hspace{0mm}+2i C^\mu(D_\mu \chi -D^\nu \chi^+ _{\mu\nu} )+\phi \{\chi,\chi\}+\frac{1}{4}\phi \{\chi^{+A},\chi^+ _A\} + \overline{\phi} \{C^\mu,C_\mu\}\nonumber  \\
& &\hspace{0mm} 
 -2V^\mu\{C^\nu, \delta_{\mu\nu}\tilde{\eta} +\chi^- _{\mu\nu}\} -2V^\mu\{\delta_{\mu\nu} \chi -\chi^+ _{\mu\nu},\tilde{\psi}^\nu\} \nonumber \\
& &+\frac{1}{4}V^\mu[\phi,[\overline{\phi},V_\mu]]+\frac{1}{4}V^\mu[\overline{\phi},[\phi,V_\mu]] \Big{)}. 
\label{eq:on-shell action}
\end{eqnarray}
This action possess the following discrete symmetry,
\begin{align}
\phi &\to -\phi,& \overline{\phi}&\to -\overline{\phi}, &\chi
 &\leftrightarrow\tilde{\eta},&
\psi^\mu &\leftrightarrow  C^\mu,& \chi^\pm_A &\to -\chi^\mp _A.
\label{eq:discrete-symmetry}
\end{align}
Applying the symmetry (\ref{eq:discrete-symmetry}) to the TSUSY
transformation, we get a new fermionic symmetry which is shown in \ref{ch:N=4 symmetry}.
Since the new symmetry satisfies the TSUSY algebra of $\{s^-\}$ part, the on-shell action (\ref{eq:on-shell action}) has the $N$=4 TSUSY
and SO(4) R-symmetry and it is equal to Marcus's $N$=4 twisted action
 \cite{Mac} when we redefine the fields.

We summarize the $N$=4 twisting procedure here\cite{LCL}.
Yamron firstly pointed out the existence of three twisting manners with
respect to $N$=4 SUSY \cite{Yam}. 
This was analyzed by Vafa and Witten \cite{VW} and Marcus
\cite{Mac,BT} in detail. 
Let us focus on the Marcus's twist. 
The Marcus's twist has two BRST charges having the same ghost number. 
The Dirac-K\"ahler twist has the same characteristic as the Marcus's
twist in the sense that there are the scalar and the pseudo-scalar BRST
charges.  
Therefore we have thought that the Dirac-K\"ahler twist has a close
relation to the one of Marcus.
We now claim that the four-dimensional Dirac-K\"ahler twist is
equivalent to the Marcus twist because we have derived the Marcus
action from our formalism.   
\setcounter{equation}{0}
\renewcommand{\theequation}{\arabic {section}.\arabic{equation}}
\section{Euclidean $N$=4 SUSY Action}

In this section, we derive an ordinary spinor type $N$=4 SUSY action using
Dirac-K\"ahler twist. The tensor fermions appearing in the twisted
theory $\{C_{\mu}, \psi_{\mu}, \chi^{+}_{A}, \chi^{-}_{A}, \chi, 
\tilde{\eta}\}$ are easily transformed into a spinor field with
the internal symmetry.
We define a Dirac-K\"ahler field as follows:
\begin{eqnarray}
\Psi_{\alpha i} = ({\bf 1} \psi + \gamma^\mu \psi_\mu 
+\frac{1}{2}\gamma^{\mu\nu}\psi_{\mu\nu} 
+\tilde{\gamma}^\mu\psi'_\mu  +\gamma_5 \tilde{\psi} )_{\alpha i}.
\end{eqnarray}
where 
$\chi = \psi+ \tilde{\psi},~ \tilde{\eta} = \psi - \tilde{\psi},~
C_\mu  = \psi_\mu + \psi'_\mu ,~ \tilde{\psi}_\mu = \psi_\mu-
 \psi'_\mu,~
\chi^{+\mu\nu} = \psi^{\mu\nu} -\frac{1}{2}\epsilon^{\mu\nu\rho\sigma}
 \psi_{\rho\sigma},~ \chi^{-\mu\nu} = -(\psi^{\mu\nu}
 +\frac{1}{2}\epsilon^{\mu\nu\rho\sigma} \psi_{\rho\sigma})$ .
Then the action (\ref{eq:on-shell action}) is rewritten as follows:
\begin{eqnarray}
S 
&=& \int d^4 x\text{Tr}\Big{(}\ \frac{1}{2}i\ \overline{\Psi}^i \gamma^\mu D_\mu \Psi^i + \frac{1}{2}\phi D^\mu D_\mu \overline{\phi}+\frac{1}{2}V^\mu D^\nu D_\nu V_\mu  + \frac{1}{2}(F^- _{\mu\nu})^2   \nonumber \\
& &\hspace{20mm}+\frac{1}{2}\phi(\overline{\Psi}^i\gamma^5 \Psi^i +\overline{\Psi}^i \Psi^j (\hat{\gamma}_5)^{ij} ) -\frac{1}{2}\overline{\phi}(\overline{\Psi}^i\gamma^5 \Psi^i -\overline{\Psi}^i \Psi^j (\hat{\gamma}_5)^{ij} ) \nonumber \\
& &\hspace{20mm}- V^\mu \overline{\Psi}^i \Psi^j (\hat{\gamma}_\mu)^{ij}  
+\frac{1}{4}V^\mu[\phi,[\overline{\phi},V_\mu]]+\frac{1}{4}V^\mu[\overline{\phi},[\phi,V_\mu]]
\nonumber \\
& &\hspace{20mm}
 -\frac{1}{4}[V_\mu,V_\nu]^2+\frac{1}{8}[\phi,\overline{\phi}]^2
\Big{)}.
\label{eq:inverse action}
\end{eqnarray}
This is an untwisted action. 
It should be noted that $V_\mu$ is not a spacetime vector field since it
couples to the internal space $\gamma$-matrix $\{(\hat{\gamma}_\mu)^{ij}\}$.  
The action (\ref{eq:inverse action}) is easily rewritten by using
$\phi^{ij}$, $\tilde{\phi}^{ij}$ as follows:
\begin{eqnarray}
S &=& \int d^4 x\text{Tr}\Big{(}\ \frac{1}{2}i\ \overline{\Psi}^i \gamma^\mu D_\mu \Psi^i + \frac{1}{8} \tilde{\phi}^{ij} D^\mu D_\mu \phi^{ji}  + \frac{1}{2}(F^- _{\mu\nu})^2   \nonumber \\
& &\hspace{15mm}+ \overline{\Psi}^i P_+ \Psi^j \phi^{ij}+ \overline{\Psi}^i P_- \Psi^j \tilde{\phi}^{ij} 
-\frac{1}{64}[\tilde{\phi}^{ij} ,\tilde{\phi}^{lm} ][\phi^{ji} ,\phi^{ml} ]
\Big{)} ,
\label{eq:N=4 action}
\end{eqnarray}
where $\phi$ and $\tilde{\phi}$ are defined as follows:
\begin{eqnarray}
\phi^{ij} &=& (\phi P_+ - \overline{\phi} P_- -V_\mu \hat{\gamma}^\mu)^{ij},
 \nonumber \\
\tilde{\phi}^{ij} &=& (\overline{\phi} P_+ - \phi P_- -V_\mu \hat{\gamma}^\mu)^{ij}.
\end{eqnarray}
$\phi^{ij}$ and $\tilde{\phi}^{ij}$ satisfy the following relations
respectively,
\begin{align}
\phi^{\dagger} &= C \phi^{T} C^{-1},& \phi^{\dagger} = \phi^{},
\nonumber \\
\tilde{\phi}^{\dagger} &= C \tilde{\phi}^{T} C^{-1},& 
\tilde{\phi}^{\dagger} = \tilde{\phi}.
\label{eq:c phi}
\end{align}
Using (\ref{eq:c phi}), we find that $C\phi$ and $C\tilde{\phi}$ are
anti-symmetric matrices. 
They are equivalently written in the following form
\begin{eqnarray}
(C \tilde{\phi})^*_{ij} = -\frac{1}{2}\epsilon_{ijkl}({C\phi})^{kl}, 
\end{eqnarray}
where $\epsilon_{1234}=1$. They are  the second rank  self-dual
tensor of the representation {\bf 6} of SU(4) group.
This untwisted theory has the internal SU(4) R-symmetry, but the
R-symmetry is reduced to SO(4) due to the Dirac-K\"ahler twist in the
twisted theory.  
This action (\ref{eq:N=4 action}) is a $N$=4 SUSY action at on-shell
level.

\renewcommand{\theequation}{\arabic {section}.\arabic{equation}}
\section{Conclusions and Discussions}
We have proposed a twisted $N$=2 superspace formalism with the central
charge based on the Dirac-K\"ahler twisting procedure. 
We have examined a general property of a superfield in this formalism. 
In this case the superfield has many superfluous fields.
We have introduced a superfield $\mathcal{V}_\mu $ and found its
constraints by means of R-symmetry. Using this superfield, we have found
the off-shell action of the twisted hypermultiplet on superspace.
It turned out that the R-symmetry played an important role in this
formalism.

We have then extended this model to the covariantized theory.
We obtained a Donaldson-Witten theory coupled to the twisted
hypermultiplet. This theory is off-shell $N$=2 TSUSY.
When the auxiliary fields are integrated out, the symmetry of the theory
is enhanced to the $N$=4 TSUSY at on-shell level. This action is 
equal to that of Marcus after some fields are redefined.

Since the Dirac-K\"ahler twisted algebra and Marcus's one include two
scalar twisted supercharges which are assigned to the same ghost
number :$+1$\cite{J-Saito} and both twists lead to the same action,  
we then claim that the four-dimensional Dirac-K\"ahler twisting
procedure is equivalent to the Marcus's twist. 
It should be noted that the Dirac-K\"ahler twist can be defined in other
dimensions \cite{KKM,J-Saito}.  
The advantage of the Dirac-K\"ahler twist is that since
all the fermions are related to spinor by the Dirac-K\"ahler mechanism,
we can easily construct a corresponding untwisted theory.  
This equivalence suggest that these $N$=2 and $N$=4 TSUSY models can be
applied to a lattice SUSY theory, because the Dirac-K\"ahler mechanism is 
compatible with a lattice theory and corresponding two-dimensional
models are realized on a lattice based on the twisted superspace
formalism \cite{DKKN,DKKN2} .

In this paper, we have considered the $N$=2 twisted superspace formalism
with a central charge.
There is an alternative approach where a theory has the $N$=4
TSUSY with a central charge and the SO(4) R-symmetry.
A corresponding untwisted theory has the maximal R-symmetry; USp(4)
$\simeq$ SO(5) which can be reduced to SO(4) by Dirac-K\"ahler twist in
twisted theory.
In this case the off-shell $N$=4 TSUSY is preserved \cite{K-Saito}.

\vspace{1cm}

\textbf{\Large Acknowledgements}\\
We would like to thank Prof. N. Kawamoto for the collaboration at an
earlier stage and instructive comments.
We would like to thank I. Kanamori, K. Nagata and  J. Saito for
fruitful discussions and comments.
In particular the comments by J. Saito are crucial in our formulation. 
We would like to thank V. de Beauce for careful reading of the manuscript. 

\newpage
\appendix
\setcounter{equation}{0}
\def\thesection{Appendix \Alph{section}}
\renewcommand{\theequation}{A.\arabic{equation}}
\section{}
\label{ch:N=4 symmetry}
We show the full list of the on-shell $N$=4 twisted SUSY transformations,

\[
 \begin{array}{|c||c|c|c|}
\hline
 &\mbox{gh}\sharp &s^+ & s^-  \\
\hline
\phi & 2& 0 &  0 \\
\overline{\phi} & -2& -2\chi &  2\tilde{\eta}    \\ 
C_\nu & 1 & \frac{i}{2}D_\nu \phi &-\frac{1}{2}[\phi,V_\nu]  \\
\tilde{\psi}_\nu & 1 &\frac{1}{2}[\phi,V_\nu]  & -\frac{i}{2}D_\nu \phi  \\
\chi^+ _B & -1  & -\frac{1}{4}\delta^+ _{B,\rho\sigma}[V^\rho,V^\sigma]- iF^+ _B & -\frac{i}{2}\delta^+_{B,\rho\sigma}D^\rho V^{\sigma}  \\
\chi^- _B & -1&\frac{i}{2}\delta^-_{B,\rho\sigma}D^\rho V^{\sigma} 
&\frac{1}{4}\delta^- _{B,\rho\sigma}[V^\rho,V^\sigma]+ iF^- _B   \\
\chi & -1 &-\frac{1}{4} [\phi,\overline{\phi}] &-\frac{i}{2} D^\mu V_\mu   \\ 
\tilde{\eta} & -1 &-\frac{i}{2} D^\mu V_\mu & -\frac{1}{4} [\phi,\overline{\phi}] \\  
\omega_\nu &0 &- C_\nu & -\tilde{\psi}_\nu  \\
V_\nu &0  &\tilde{\psi}_\nu & C_\nu   \\
\hline
\end{array}
\]

\[
 \begin{array}{|c||c|c|}
\hline
 &s^+ _\mu & s^- _\mu \\
\hline
\phi &-2C_\mu  &  2\tilde{\psi}_\mu \\
\overline{\phi} & 0 & 0 \\ 
C_\nu &-\frac{1}{4}\delta^+ _{\mu\nu,\rho\sigma}[V^\rho,V^\sigma] +i F^- _{\mu\nu} +\frac{1}{4}\delta_{\mu\nu} [\phi,\overline{\phi}] & \frac{i}{2} (\delta^+ _{\mu\nu,\rho\sigma}D^\rho V^\sigma + \delta_{\mu\nu} D^\rho V_{\rho} -2D_{\mu} V_{\nu})  \\
\tilde{\psi}_\nu &\frac{i}{2} (\delta^- _{\mu\nu,\rho\sigma}D^\rho V^\sigma + \delta_{\mu\nu} D^\rho V_{\rho} -2D_{\mu} V_{\nu})  &  -\frac{1}{4}\delta^- _{\mu\nu,\rho\sigma}[V^\rho,V^\sigma] +i F^+ _{\mu\nu} +\frac{1}{4}\delta_{\mu\nu} [\phi,\overline{\phi}] \\
\chi^+ _B &-\frac{i}{2}\delta^+ _{B,\mu\nu}D^\nu \overline{\phi} &- \frac{1}{2}\delta^+_{B,\mu\nu}[\overline{\phi},V^\nu] \\
\chi^- _B & -\frac{1}{2}\delta^-_{B,\mu\nu}[\overline{\phi},V^\nu] & -\frac{i}{2}\delta^- _{B,\mu\nu}D^\nu \overline{\phi} \\
\chi &\frac{i}{2}D_\mu \overline{\phi} & -\frac{1}{2}[\overline{\phi},V_\mu]\\ 
\tilde{\eta} &\frac{1}{2}[\overline{\phi},V_\mu] & -\frac{i}{2}D_\mu \overline{\phi} \\
\omega_\nu &- (\delta_{\mu\nu}\chi-\chi^+ _{\mu\nu} ) & -(\delta_{\mu\nu} \tilde{\eta}+\chi^- _{\mu\nu} )\\
V_\nu &\delta_{\mu\nu} \tilde{\eta}-\chi^- _{\mu\nu} & \delta_{\mu\nu}\chi+\chi^+ _{\mu\nu} \\

\hline
\end{array}
\]

\[\hspace{-13mm}
 \begin{array}{|c||c|c|}
\hline
 & s^+ _A & s^- _A\\
\hline
\phi  & 0  &0 \\
\overline{\phi} & -2\chi^{+}_A  & -2\chi^{-}_A\\ 
C_\nu & -\frac{i}{2}\delta^+ _{A,\nu\rho}D^\rho \phi  & -\frac{1}{2}\delta^- _{A,\mu\nu}[\phi,V^\nu] \\
\tilde{\psi}_\nu &    \frac{1}{2}\delta^+ _{A,\mu\nu}[\phi,V^\nu]&
\frac{i}{2}\delta^- _{A,\nu\rho}D^\rho \phi \\
\chi^+ _B & \frac{i}{8} \Gamma^+ _{ABC} (\frac{i}{2}\delta^{+C,\rho\sigma}[V_\rho,V_\sigma]+2F^{+C}) -\frac{1}{4} \delta^+ _{A,B}[\phi,\overline{\phi}] &  \frac{i}{2}\delta^- _{A,\rho\nu}\delta^+ _{B,\sigma}{}^\nu D^\rho V^\sigma  \\
\chi^- _B & -\frac{i}{2}\delta^+ _{A,\rho\nu}\delta^- _{B,\sigma}{}^\nu D^\rho V^\sigma  & \frac{i}{8} \Gamma^- _{ABC} (-\frac{i}{2}\delta^{-C,\rho\sigma}[V_\rho,V_\sigma]-2F^{-C}) +\frac{1}{4} \delta^- _{A,B}[\phi,\overline{\phi}] \\
\chi & \frac{1}{4}\delta^+ _{A,\rho\sigma}[V^\rho,V^\sigma]+i F^+ _A & -\frac{i}{2}\delta^- _{A,\rho\sigma}D^\rho V^\sigma \\ 
\tilde{\eta} &-\frac{i}{2}\delta^+ _{A,\rho\sigma}D^\rho V^\sigma & \frac{1}{4}\delta^- _{A,\rho\sigma}[V^\rho,V^\sigma]+i F^- _A \\ 
\omega_\nu & - \delta^+ _{A,\nu\rho} C^\rho & - \delta^- _{A,\nu\rho} \tilde{\psi}^\rho\\
V_\nu  &
-\delta^+ _{A,\nu\rho}\tilde{\psi}^\rho & -\delta^- _{A,\nu\rho} C^\rho\\
\hline
\end{array}
\]

\[
 \begin{array}{|c||c|c|}
\hline
 & R^+ _A & R^- _A\\
\hline
\phi  & 0  &0 \\
\overline{\phi} & 0  & 0 \\ 
C_\nu & -\frac{i}{2}\delta^+ _{A,\nu\rho}C^\rho  &0 \\
\tilde{\psi}_\nu & 0 &
- \frac{i}{2}\delta^- _{A,\nu\rho}\tilde{\psi}^\rho \\
\chi^+ _B & \frac{i}{2} \delta^+ _{A,B}\chi -\frac{i}{8}\Gamma^+ _{ABC}\chi^{+C} & 0  \\
\chi^- _B & 0  &   -\frac{i}{2} \delta^- _{A,B}\tilde{\eta} -\frac{i}{8}\Gamma^- _{ABC}\chi^{-C} \\
\chi &- \frac{i}{2}\chi^+ _A &0 \\ 
\tilde{\eta} & 0 & \frac{i}{2}\chi^- _A \\ 
\omega_\nu & 0 &  0 \\
V_\nu  &
-\frac{i}{2}\delta^+ _{A,\nu\rho}V^\rho  & -\frac{i}{2}\delta^- _{A,\nu\rho}V^\rho \\
\hline
\end{array}
\]

The $N$=4 twisted supercharges $\{s^{\pm}, s^{\pm}_{\mu}, s^{\pm}_{A}\}$
satisfy the following commutation relations up to gauge
transformation at on-shell level. 
\begin{eqnarray}
\{s^{\pm},s^{\pm}\}\varphi &=& \delta_{gauge(\mp i\phi)}\varphi, \hspace{2cm}
\{s^{\pm},s^{\pm}_{\mu}\}\varphi = -i(\partial_{\mu}\varphi +
\delta_{gauge(-\omega_{\mu})}\varphi), \nonumber\\
\{s^{\pm}_{\mu},s^{\pm}_{\nu}\}\varphi &=& \delta_{\mu\nu}
\delta_{gauge(\mp i\bar{\phi})}\varphi, \hspace{1.5cm}
\{s^{\pm}_{\mu},s^{\pm}_{A}\}\varphi = i\delta^{\pm}_{A,\mu\rho}
(\partial^{\rho}\varphi+\delta_{gauge(-\omega^{\rho})}\varphi), \nonumber\\
\{s^{\pm}_{A},s^{\pm}_{B}\}\varphi &=& \delta^{\pm}_{A,B}
\delta_{gauge(\mp i\phi)}\varphi, \hspace{1.3cm}
\{s^{\pm},s^{\pm}_{A}\}\varphi = 0, \nonumber \\
\{s^\pm,s^\mp _\mu\}\varphi &=& \delta_{gauge(iV_\mu)}\varphi , \hspace{21mm} 
 \{s^\pm,s^\mp _A\}\varphi = 0, \nonumber \\ 
 \{s^\pm _\mu ,s^\mp _A\}\varphi &=& \delta^\mp _{A,\mu\nu}\delta_{gauge(iV^\nu)}\varphi,
 \hspace{12mm} \{s^+,s^-\}\varphi = 0, \nonumber \\
\{s^+_\mu, s^- _\nu\}\varphi &=& 0, \hspace{38mm} \{s^+ _A,s^- _B\}\varphi=0,
\end{eqnarray}
where 
$\delta_{gauge(\varepsilon)}\omega_{\mu}=D_{\mu}\varepsilon$,~ 
$\delta_{gauge(\varepsilon)}\varphi' = -i[\varphi', \varepsilon]$,~ 
$\varphi' = \{\phi, \bar{\phi}, C_\nu, \tilde{\psi}_\nu,
\chi^+_B, \chi^-_B, \chi, \tilde{\eta}, V_\nu \}$. 

\setcounter{equation}{0}
\renewcommand{\theequation}{B.\arabic{equation}}
\section{}
We define an Euclidean four dimensional $\gamma$-matrices:
\begin{eqnarray}
\{\gamma_\mu, \gamma_\nu \} = 2\delta_{\mu\nu}, \qquad 
\gamma^{\mu\dagger} = \gamma^\mu,
\end{eqnarray}
where $\gamma_\mu$  satisfies the Clifford algebra.
 We introduce the following notations:
\begin{eqnarray}
\gamma_{\mu\nu}\equiv \frac{1}{2}[\gamma_\mu, \gamma_\nu],\qquad
\tilde{\gamma}_\mu\equiv \gamma_\mu\gamma_5.
\end{eqnarray}
We use the following representation of $\gamma$-matrix:
\begin{eqnarray}
\gamma^\mu =
\left(\begin{array}{cc}
0 & i\sigma^\mu \\
i\bar{\sigma}^\mu & 0
\end{array}\right),
\end{eqnarray}
where $\sigma^\mu = (\sigma^1, \sigma^2, \sigma^3, \sigma^4)$,
$\bar{\sigma}^\mu = (-\sigma^1, -\sigma^2, -\sigma^3, \sigma^4)$ with
$\sigma^4 = -i {\bf 1}_{2\times 2}$ and $\sigma^i$ are Pauli matrices, 
$i\in \{1,2,3\}$. A charge conjugation matrix $C$ and $B$ matrix are in
general defined as follows:
\begin{align}
\gamma_{\mu} &= \eta B^{-1}\gamma_{\mu}^* B ,& B^* B &= \epsilon\bf{1},
  \nonumber\\
\gamma_\mu &= \eta C^{-1}\gamma_\mu ^T C,& C^T &= \epsilon C,
\end{align}
where $(\eta,\epsilon)=(\pm1,-1) $.

In a four dimensional Euclidean space Majorana fermions do not exist
because the factor $\epsilon$ should be equal to $-1$ \cite{Kugo-T}. A Majorana spinor satisfies
the following condition,
\begin{eqnarray}
\psi^* = B\psi,
\label{eq:def majorana}
\end{eqnarray}
which leads
\begin{eqnarray}
\psi = B^*\psi = B^* B \psi.
\end{eqnarray}
Thus the existence of Majorana fermion requires $B^*B=1$.
This condition can not be taken in four dimensional Euclidean space.
We can, however, take a SU(2)$\simeq$USp(2) Majorana fermion and
a USp(4) Majorana fermion which satisfies the following condition,
respectively,  
\begin{eqnarray}
\psi^{i*} &=& \epsilon^{ij} B\psi^j, 
\label{eq:majorana} \\
\psi^{l*} &=& C^{lm} B\psi^m, 
\end{eqnarray}
where $i,j\in \{1,2\},\quad  l,m \in\{ 1,2,3,4\}$ and these fermions
correspond to the fermions which appear in $N$=2 and $N$=4 supersymmetric
theory, respectively. 
In this paper we choose $(\eta,\epsilon)=(1,-1)$ and $C=B=-\gamma_1 \gamma_3$.

\setcounter{equation}{0}
\renewcommand{\theequation}{C.\arabic{equation}}
\section{}
\label{ch:algebra}
In this appendix we explain SUSY algebra (\ref{eq:SUSY al. with central
charge}) with respect to two-component spinors.
For simplicity we consider Minkowski space. We represent the
supercharge appearing in algebra (\ref{eq:SUSY al. with central
charge}) as the following form,
\begin{eqnarray}
Q^{i}_{\alpha}&=& 
 \left(
    \begin{array}{c}
     -i\epsilon^{ij} \mathsf{Q}_{\alpha j} \\
      \overline{\mathsf{Q}}^{\dot{\alpha}i} 
    \end{array}
\right),\\
\overline{Q}_{\alpha i} &=& (\mathsf{Q}_i ^\alpha , i \epsilon_{ij}\overline{\mathsf{Q}}^j _{\dot{\alpha}}).
\end{eqnarray}
In this notation the SU(2) Majorana condition (\ref{eq:majorana}) is the
following form, 
\begin{eqnarray}
\overline{\mathsf{Q}}^i _{\dot{\alpha}} = (\mathsf{Q}_{\alpha i})^\dagger.
\end{eqnarray}
We describe the algebra (\ref{eq:SUSY al. with central charge}) with
respect to theses two-component supercharges,
\begin{align}
\{ \mathsf{Q}_{\alpha i}, \overline{\mathsf{Q}}^j_{\dot{\alpha}}\} &=2\delta_i ^j
 (\sigma^\mu)_{\alpha\dot{\alpha}} P_\mu,& \{\mathsf{Q}_{\alpha i}, \mathsf{Q}_{\beta j}\}
 = 2i\epsilon_{ij}\epsilon_{\alpha\beta}Z,
\end{align}
where supercharges with upper and lower indices are related through the $\epsilon$-tensor. 

\setcounter{equation}{0}
\renewcommand{\theequation}{D.\arabic{equation}}
\section{}
We introduce the definition of $\delta^\pm_{A, B}$ and 
$\Gamma^\pm_{ABC}$. The suffix $A$ is the second rank tensor
which denotes the suffix $\mu\nu$: $\mu, \nu \in \{ 1, \cdots,
4$ \}. The definition of $\delta^\pm_{A,B}$ is 
\begin{eqnarray}
\delta^\pm_{A,B} = \delta^{\pm}_{\mu\nu, \rho\sigma}
= \delta_{\mu\rho}\delta_{\nu\sigma}
 -\delta_{\mu\sigma}\delta_{\nu\rho} \mp \epsilon_{\mu\nu\rho\sigma},
\end{eqnarray}
where $\delta^{\pm A,B}\delta^{\mp}_{~A,B}=0$.
(Anti-)self-dual tensors $\chi^{\pm A}$ satisfy 
\begin{eqnarray}
\chi^{\pm A} = \frac{1}{4}\delta^{\pm A,B}\chi_{B},
\end{eqnarray}
where $\chi_{B}=\chi^+_{B} +\chi^-_{B}$. 

Variants of the definition of $\Gamma_{ABC}$ which stand for the third
anti-symmetric tensor for $ABC$ is 
\begin{eqnarray}
\Gamma^{\pm \mu\alpha, \nu\beta, \rho\gamma} &=& \delta^{\alpha\nu}
\delta^{\beta\rho}\delta^{\gamma\mu} +\delta^{\mu\nu}\delta^{\beta\gamma}
\delta^{\rho\alpha} + \delta^{\alpha\beta}\delta^{\nu\gamma}\delta^{\rho\mu}
+\delta^{\mu\beta}\delta^{\nu\rho}\delta^{\gamma\alpha} \nonumber\\
&& -(\delta^{\alpha\nu}\delta^{\beta\gamma}\delta^{\rho\mu} +\delta^{\mu\nu}
\delta^{\beta\rho}\delta^{\gamma\alpha} +\delta^{\alpha\beta}\delta^{\nu\rho}
\delta^{\gamma\mu} +\delta^{\mu\beta}\delta^{\nu\gamma}\delta^{\rho\alpha})
\nonumber \\
&& \mp\epsilon^{\mu\alpha\beta\gamma}\delta^{\beta\gamma} \mp
\epsilon^{\mu\alpha\nu\rho}\delta^{\beta\gamma} \pm 
\epsilon^{\mu\alpha\nu\gamma}\delta^{\beta\rho} \pm
\epsilon^{\mu\alpha\beta\rho}\delta^{\nu\gamma} \nonumber \\
&=& \delta^{\pm\mu\alpha,\nu\rho}\delta^{\beta\gamma} +\delta^{\pm\mu\alpha,
\beta\gamma}\delta^{\nu\rho} -\delta^{\pm\mu\alpha,\nu\gamma}
\delta^{\beta\rho} -\delta^{\pm\mu\alpha,\beta\rho}\delta^{\nu\gamma}, \\
\Gamma^{\pm AB\rho\sigma} &=& 
\frac{1}{2}(\delta^{\pm A,\nu\rho}\delta^{\pm B},_{\nu}^{~\sigma}
-\delta^{\pm A, \nu\sigma}\delta^{\pm B},_{\nu}^{~\sigma}), \\
\Gamma^{\pm ABC} &=& \frac{1}{4}\delta^{\pm A,\nu\rho}
\delta^{\pm B},_{\nu}^{~\sigma}\delta^{\pm C},_{\rho\sigma}.
\end{eqnarray}



\begin{thebibliography}{30}

\bibitem{W}
 E. Witten, 
 Commun. Math. Phys. \textbf{117} (1988) 353; 
 Commun. Math. Phys. {\bf 118} (1988) 411. 




 \bibitem{bs}
  L. Baulieu and I. Singer,
  Nucl. Phys. Proc. Suppl. {\bf B5} (1988) 12.

 \bibitem{bms}
  R. Brooks, D. Montano and J. Sonnenschein, 
  Phys. Lett. {\bf B214} (1988) 91.

 \bibitem{lp}
  J.M.F. Labastida and M. Pernici, 
  Phys. Lett. {\bf B212} (1988) 56; 
  Phys. Lett. {\bf B213} (1988) 319.

\bibitem{BRT1}
 D. Birmingham, M. Rakowski and G. Thompson, 
 Phys. Lett. {\bf B214} (1988) 381; 
 Phys. Lett. {\bf B212} (1988) 187; 
 Nucl. Phys. {\bf B315} (1989) 577.



\bibitem{BR}
 D. Birmingham and M. Rakowski, 
 Mod. Phys. Lett. \textbf{A4}(1989) 1753; Phys. Lett. \textbf{B269} 
 (1991) 103; Phys. Rev. \textbf{B272} (1991) 217.

\bibitem{BRT}
 D. Birmingham, M. Rakowski, and G. Thompson, Nucl. Phys. \textbf{B329},
(1990) 83.








\bibitem{DGS}
 F. Delduc, F. Gieres, and S.P.Sorella,
	Phys. Lett. {\bf B225} (1989) 367.


\bibitem{GMS}
 E. Guadagnini, N. Maggiore and S.P. Sorella, 
 Phys. Lett. {\bf B247} (1990) 543;
 Phys. Lett. {\bf B255} (1991) 65.

\bibitem{BM}
A. Blasi and N. Maggiore, 
Class. Quantum Grav. {\bf 10} (1993) 37.

\bibitem{MSore}
 N. Maggiore and S.P. Sorella,
 Int. Journ. Mod. Phys. {\bf A8} (1993) 325.

\bibitem{LPS}
 C. Lucchesi, O. Piguet and S. P. Sorella,
 Nucl. Phys. {\bf B395} (1993) 325.\\

\bibitem{FTVVSS}
 F. Fucito, A. Tanzini, L.C.Q. Vilar, O.S. Ventura, C.A.G. Sasaki, 
 S.P. Sorella, 
 hep-th/9707209.


\bibitem{LSZ}
 R. Leitgeb, M. Schweda, and H. Zerrouki, 
 Nucl. Phys \textbf{B542} (1999) 425.

\bibitem{GGPS}
 F. Gieres, J. Grimstrup. T. Pisar and M. Schweda, JHEP {\bf 0006} (2000) 018.


\bibitem{BSSV}
J. L. Bold, C. A. G. Sasaki, S. P. Sorella, and L. C. Q. Vilar, 
J. Phys. \textbf{A34} (2001) 2743. 


\bibitem{LSSTV}
 V.E.R. Lemes, M.S. Sarandy, S.P. Sorella, A. Tanzini and O.S. Ventura, 
 JHEP {\bf 0101} (2001) 016 .

\bibitem{GGNPS}
F. Gieres, J. Grimstrup, H. Nieder, T. Pisar and M. Schweda,
Phys. Rev. {\bf D66} (2002) 025027




\bibitem{LL}
J.M.F. Labastida and P.M. Llatas, 
Nucl. Phys. {\bf B379} (1992) 220. 

\bibitem{CLS}
 O. M. Del Cima, K. Landsteiner, and M. Schweda, 
Phys.Lett. {\bf B439} (1998) 289.



   


\bibitem{KT}
 N. Kawamoto and T. Tsukioka, Phys. Rev. \textbf{D61}(2000)105009.



\bibitem{KW}
 N. Kawamoto and Y. Watabiki, Commun. Math. Phys. \textbf{144}, 641(1992);\\
 Mod. Phys. Lett. {\bf A7} (1992) 1137.

\bibitem{KW2}
 N. Kawamoto and Y. Watabiki, Phys.Rev.\textbf{D45} (1992) 605;\\
 Nucl. Phys. {\bf B396} (1993) 326.


 \bibitem{KOS}
  N. Kawamoto, E. Ozawa and K. Suehiro,
  Mod. Phys. Lett. {\bf A12} (1997) 219.

\bibitem{KSTU}
  N. Kawamoto, K. Suehiro, T. Tsukioka and H. Umetsu, 
  Commun. Math. Phys. {\bf 195} (1998) 233, 
  Nucl. Phys. {\bf B532} (1998) 429.



\bibitem{IL} 
D. Ivanenko and L. Landau,  Z. Phys. \textbf{48} (1928) 340.

\bibitem{Kahler}
 E. K\"ahler, Rend. Math. Appl. \textbf{21} (1962) 425.


 \bibitem{G}
  W. Graf, 
  Ann. Inst. Henri Poincare {\bf A29} (1978) 85.

 \bibitem{BJ}
  P. Becher and H. Joos,
  Z. Phys. {\bf C15} (1982) 343. 

 \bibitem{Rabin}
  J.M. Rabin, 
  Nucl. Phys. {\bf B201} (1982) 315.

 \bibitem{BDH}
  T. Banks, Y. Dothan and D. Horn,
  Phys. Lett. {\bf B117} (1982) 413. 

 \bibitem{BennT}
  I.M. Benn and R.W. Tucker, 
  Commun. Math. Phys. {\bf 89} (1983) 341. 

 \bibitem{Bull}
  J.A. Bullinaria, 
  Ann. Phys. {\bf 168} (1986) 301. 


\bibitem{DVF}
  P. Di Vecchia and S. Ferrara, 
  Nucl. Phys. {\bf B130} (1977) 93. 

 \bibitem{SchapT}
  F.S. Schaposnik and G. Thompson, 
  Phys. Lett. {\bf B224} (1989) 379.




\bibitem{KKU}
J. Kato, N. Kawamoto and Y. Uchida,
 Int. Jour. Mod. Phys. {\bf A 19} (2004) 2149.



\bibitem{KKM}
J. Kato, N. Kawamoto, A. Miyake,
 Nucl. Phys. {\bf B721} (2005) 229. 



\bibitem{GSW}
 R. Grim, M. Sohnius and J. Wess, 
 Nucl. Phys. {\bf B 133} (1978) 275. 

\bibitem{Sohnius}
 M. Sohnius, 
 Nucl. Phys. {\bf B 136} (1978) 461.


\bibitem{Fay}
 P. Fayet
Nucl. Phys. {\bf B113} (1976) 135.

\bibitem{Soh}
 M.F. Sohnius.  Nucl .Phys. {\bf B138} (1978) 109.






\bibitem{AL}
M. Alvarez and J.M.F. Labastida, 
Nucl. Phys. {\bf B437} (1995) 356.

\bibitem{Mac}
N. Marcus,
Nucl. Phys. {\bf B452} (1995) 331.

\bibitem{Harmonic}
A.~Galperin, E.~Ivanov, V.~Ogievetsky and E.~Sokatchev,
``Harmonic Superspace'', Cambridge University Press , 2001


\bibitem{KSuss}
  J. Kogut and L. Susskind, 
  Phys. Rev. {\bf D11} (1975) 395.

\bibitem{Suss}
  L. Susskind, Phys. Rev. {\bf D16} (1977) 3031.

\bibitem{KS} 
  N. Kawamoto and J. Smit, 
  Nucl. Phys. {\bf B192} (1981) 100. 



\bibitem{DKKN} 
 A. D'Adda, I. Kanamori, N. Kawamoto and K. Nagata, 
 Nucl. Phys. {\bf B707} (2005) 100; 
hep-lat/0507029 (to be published in Phys. Lett. {\bf B}).

\bibitem{DKKN2} 
 A. D'Adda, I. Kanamori, N. Kawamoto and K. Nagata, in preparation.

\bibitem{C}
S. Catterall, JHEP {\bf 0411} (2004) 006;
JHEP {\bf 0506} (2005) 027.


\bibitem{C2}
S. Catterall, Phys. Lett. {\bf B549} (2002) 253. 


\bibitem{S}
F. Sugino, JHEP {\bf 0403} (2004) 067;  JHEP {\bf 0401} (2004) 015;
 JHEP {\bf 0501} (2005) 016.


\bibitem{LCL}
J. M. F. Labastida and C. Lozano,
Nucl. Phys. {\bf B502} (1997) 741.



\bibitem{Yam}
 J.P. Yamron, 
 Phys. Lett. {\bf B213} (1988) 325.



\bibitem{VW}
C. Vafa and E. Witten,
Nucl. Phys. {\bf B431} (1994) 3.

\bibitem{BT}
 M. Blau and G. Thompson, 
 Ann. Phys. (N.Y.) {\bf 205} (1991) 130.






\bibitem{SSW}
 M. Sohnius, K.S. Stelle and P.C. West, 
 Phys. Lett {\bf 92 B} (1980) 123; 
 Nucl. Phys. {\bf B 173} (1980) 127. 

\bibitem{J-Saito}
J. Saito, Soryushironkenkyu(kyoto), {\bf 111}(2005) 117, hep-th/0512226.

\bibitem{K-Saito}
N. Kawamoto and J. Saito, to appear.

\bibitem{Kugo-T}
T. Kugo and  P. Townsend, 
 Nucl. Phs. {\bf B221} (1983) 357.




\end{thebibliography}
\end{document}